\documentstyle[12pt]{article}

\begin{document}
\def\be{\begin{equation}}
\def\ee{\end{equation}}
\def\bea{\begin{eqnarray}}
\def\eea{\end{eqnarray}}
\def\nn{\nonumber \\}
\def\e{{\rm e}}

\thispagestyle{empty}
\renewcommand{\thefootnote}{\fnsymbol{footnote}}

\  \hfill
\begin{minipage}{3.5cm}
YITP-01-74\\
November 2001 \\
hep-th/0111008 \\
\end{minipage}
%\vfill
\vspace*{2cm}

\begin{center}
{\Large\bf Holographic Entropy on the Brane}\\ 
\vspace*{0.1cm}
{\Large \bf in de Sitter Schwarzschild Space}

\vspace*{1.5cm}
%\vfill

{\sc Sachiko OGUSHI}\footnote{Electronic address: 
ogushi@yukawa.kyoto-u.ac.jp}\\

\vspace*{1cm}

{\sl Yukawa Institute for Theoretical Physics, \\
Kyoto University, Kyoto 606-8502, JAPAN}

\vspace*{2.0cm}

{\bf ABSTRACT}
\end{center}
The relationship between the entropy of de Sitter (dS) 
Schwarzschild space and that of the CFT, which
lives on the brane, is discussed by using 
Friedmann-Robertson-Walker (FRW) equations and 
Cardy-Verlinde formula.  
The cosmological constant appears on the brane 
with time-like metric in dS Schwarzschild background.  
On the other hand, in case of the brane with 
space-like metric in dS Schwarzschild background, 
the cosmological constant of the brane
does not appear because we can choose brane tension 
to cancel it.  We show that when the brane crosses 
the horizon of dS Schwarzschild black hole, 
both for time-like and space-like cases, the entropy of 
the CFT exactly agrees with the black hole entropy 
of 5-dimensional dS background 
as it happens in the AdS/CFT correspondence.

\newpage
\setcounter{page}{1}
\renewcommand{\thefootnote}{\arabic{footnote}}
\addtocounter{footnote}{-1}

\section{Introduction}

The holographic duality which connects $d+1$-dimensional 
gravity in Anti-de Sitter (AdS) background with 
$d$-dimensional conformal field theory (CFT) 
has been discussed vigorously for some years\cite{AdS}.  
The one of the evidences for the existence of the AdS/CFT 
correspondence is that the isometry of $d+1$-dimensional AdS space 
$SO(d,2)$ is identical with the conformal symmetry of 
$d$-dimensional Minkowski space.  
Recently much attention has been paid 
for the duality between de Sitter (dS) gravity and CFT 
by the analogy of the AdS/CFT correspondence\cite{AS,dS,CI}, because  
the isometry of $d+1$-dimensional de Sitter space, $SO(d+1,1)$, 
exactly agrees with the conformal symmetry of 
$d$-dimensional Euclidean space.  
Thus it might be natural to expect the correspondence 
between $d+1$-dimensional gravity in de Sitter space and 
$d$-dimensional Euclidean CFT (the dS/CFT correspondence).  
Moreover the holographic principle 
between the radiation dominated Friedmann-Robertson-Walker (FRW) 
universe in $d$-dimensions and same dimensional CFT with a dual
$d+1$-dimensional AdS description was studied in ref.\cite{EV}.  
Especially, we can see the correspondence between
black hole entropy and the entropy of the CFT
which is derived by making the appropriate
identifications for FRW equation with
the generalized Cardy formula\footnote{
The Cardy formula\cite{CD} is originally the entropy 
formula of the CFT only for 2-dimensions.}.  
The generalized Cardy formula expresses the entropy formula 
of the CFT in any dimensions\cite{EV}.  
From the point of brane-world physics\cite{RS},
the CFT/FRW relation sheds further light on the study
of the brane CFT in the background of AdS Schwarzschild 
black hole\cite{SV}.  There was much activity on the studies
of related questions\cite{others,SO,SD,DF,DF1}.  

The purpose of this letter is the further study of the CFT
in de Sitter (dS) Schwarzschild background 
guided by the analogy of AdS Schwarzschild background.  
The investigation of dS brane in dS Schwarzschild 
background in terms of FRW equations has been initiated in 
ref.\cite{SD}.
The important difference between AdS space and dS space 
is the sign of cosmological constant.  
In case of the brane with time-like (Minkowski) metric 
on AdS Schwarzschild background\cite{SV},   
the cosmological constant does not appear because
we can choose brane tension to cancel it.  But 
it is impossible for dS Schwarzschild background 
with the positive cosmological constant.  
We will see that the cosmological constant always appears in 
FRW equations deduced from time-like brane trajectory in
dS Schwarzschild background
\footnote{The work deriving brane cosmological equation
in a systematic way for a brane embedded in a bulk
with a cosmological constant has been first examined in ref.\cite{DF}.}.  
It is interesting to note that the brane with space-like (Euclidean) metric 
in dS Schwarzschild background, the cosmological constant 
of the brane does not appear for the same reason in case of AdS 
Schwarzschild background.
From the point of view of the dS/CFT correspondence, 
the investigation of space-like brane is more 
interesting than that of time-like brane.  

Furthermore we argue the entropy of the brane CFT which
is derived by using generalized Cardy formula for both 
time-like and space-like branes.  
We will see that when the brane crosses the horizon of dS Schwarzschild 
black hole, both for time-like and space-like branes, the entropy of 
the CFT is identical with the black hole entropy 
of 5-dimensional dS background as it happens in the
AdS/CFT correspondence.

\section{FRW equations in the background of de Sitter\\
Schwarzschild black hole}

We first consider a 4-dimensional time-like 
brane in 5-dimensional dS Schwarzschild background.  
From the analogy of the AdS/CFT correspondence, 
we can regard that 4-dimensional CFT exists on the brane 
which is the boundary of the 5-dimensional 
dS Schwarzschild background.  
The dynamics of the brane is described by the boundary action:
\bea
{\cal L}_{b} ={-1 \over 8\pi G_{5}}\int_{\partial {\cal M}}
\sqrt{-g}{\cal K}+{\kappa \over 8\pi G_{5}}
\int_{\partial {\cal M}} \sqrt{-g}\ , \quad {\cal K}={\cal K}^{i}_{i}
\eea
Here $G_5$ is 5-dimensional bulk Newton constant, 
$\partial {\cal M}$ denotes the surface of the brane, $g$ is the 
determinant of the induced metric on $\partial {\cal M}$, 
${\cal K}_{ij}$ is the extrinsic curvature, 
$\kappa$ is a parameter related to tension of the brane.  

From this Lagrangian, we can get the equation 
of motion of the brane as follows\cite{SV}:
\bea
\label{eom}
{\cal K}_{ij}={\kappa \over 2}g_{ij}\ ,
\eea
which implies that $\partial {\cal M}$ is a brane of 
constant extrinsic curvature.  
The bulk action is given by 5-dimensional Einstein
action with cosmological constant.
The dS Schwarzschild space is one of the exact solutions
of bulk equations of motion and can be written in the
following form,
\bea
\label{SAdS}
ds^{2}_{5}&=&\hat G_{\mu\nu}dx^\mu dx^\nu \nn
&=& -\e^{2\rho} dt^2 + \e^{-2\rho} dr^2 
+ r^2 d \Omega_{3}^2 \ ,\nn
\e^{2\rho}&=&{1 \over r^{2} }\left( -\mu + r^{2}  
- {r^4 \over l^2} \right) \ .
\eea
Here $l$ is the curvature radius of dS and $\mu$ is the
black hole mass.  In case of AdS Schwarzschild 
gravity, there is a holographic relation
between FRW brane universe which is reduction from AdS Schwarzschild 
background and boundary CFT which lives on the brane\cite{SV,SO}.  
We assume that there are some holographic relations 
between FRW universe which is reduction from dS Schwarzschild
background and boundary CFT \footnote{The investigation of dS brane 
in dS Schwarzschild background in terms of FRW equations 
has been initiated in ref.\cite{SD}.}. 
To investigate it, we rewrite dS Schwarzschild metric 
(\ref{SAdS}) in the form of FRW metric by using a new time 
parameter $\tau$ \footnote{The method discussed here
is based on the work\cite{SV}.}.  
And the parameter $t$ and $r$ in (\ref{SAdS}) are the function 
of $\tau$, namely $r=r(\tau), t=t(\tau)$.  For the purpose of 
getting the 4-dimensional FRW metric, we impose the following 
condition,
\bea
\label{cd1}
-e^{2\rho}\left( {\partial t \over \partial \tau} \right)^2
+e^{-2\rho}\left( {\partial r \over \partial \tau} \right)^2
= -1 \ .
\eea
Thus we obtain FRW metric:
\bea
\label{met1}
ds^{2}_{4}=g_{ij} dx^i dx^j = -d\tau ^2 +r^2 d \Omega_{3}^2 \ .
\eea
The extrinsic curvature, ${\cal K}_{ij}$, of the brane
can be calculated and expressed in term of the function
$r(\tau)$ and $t(\tau)$.  Thus one rewrites
the equations of motion (\ref{eom}) as
\bea
\label{cd2}
{dt \over d\tau}= -{\kappa r  \over 2}e^{-2\rho} \ .
\eea
Using (\ref{cd1}) and (\ref{cd2}), we can derive 
FRW equation for a radiation dominated universe,
Hubble parameter $H$ which is defined by $H={1\over r}{dr \over d\tau}$
is given by 
\bea
\label{HH}
H^2= {1 \over l^2} - {1 \over r^2} + {\mu \over r^4} + {\kappa^2 \over 4}\ .
\eea
Following AdS Schwarzschild gravity case\cite{SV}, 
we choose $\kappa ={2/l}$ from now on \footnote{
From the point of view of brane-world physics\cite{RS}, 
the tension of brane should be determined without ambiguity.
In fact, we  can calculate it to cancel the leading divergence
of bulk AdS Schwarzschild\cite{SO}.}.  
This equation can be rewritten by
using 4-dimensional energy $E_4$ and volume $V$ 
in the form of the standard FRW equation with
the positive cosmological constant $\Lambda$:
\bea
\label{F1}
H^2 &=& - {1 \over r^2} + {8\pi G_4 \over 3}
{E_4 \over V}+{\Lambda \over 3} \ ,\nn
E_4 &=&{ 3 \mu V  \over 8\pi G_4 r^4},\quad
\Lambda ={ 6 \over l^2 }\ .
\eea
Here $G_{4}$ is the 4-dimensional gravitational
coupling, which is defined by
\bea
\label{gg}
G_{4}={2 G_{5} \over l}\ .
\eea
$E_4$ can be regarded as 4-dimensional 
energy on the brane in dS Schwarzschild 
background which is identical with AdS 
Schwarzschild case.  The cosmological constant $\Lambda$ 
does not appear in AdS Schwarzschild background 
because we can choose brane tension $\kappa$ to cancel
the cosmological constant of AdS Schwarzschild background.
But it is impossible for dS Schwarzschild case 
because if we choose brane tension
to cancel the cosmological constant of dS Schwarzschild background, 
the brane tension should be imaginary. 

By differentiating eq.(\ref{F1}) with respect to $\tau $, 
we obtain the second FRW equation:
\bea
\label{2FR1}
\dot H &=& - 4\pi G_4 \left({E_4 \over V} 
+ p\right) + {1 \over  r^2}\ ,\nn
p &=& {\mu \over 8 \pi G_4 r^4 }\ .
\eea
Here $p$ is 4-dimensional pressure of the 
matter on the boundary.
From eqs.(\ref{F1}) and (\ref{2FR1}), we find that
the energy-momentum tensor is traceless:\footnote{
The stress-energy tensor of CFT was caldulated
in some asymptotically dS space in a sense
of dS/CFT correspondence\cite{CI}.}
\be
\label{trace2}
{T^{{\rm matter}\ \mu}}_\mu=-{E_4 \over V} + 3p 
= 0\ .
\ee
Therefore the matter on the brane can be regarded as
the radiation, which is consistent with ref.\cite{CI}.  
This means the field theory on the brane should be 
CFT as in case of AdS Schwarzschild background\cite{SV}.  

Next, we consider space-like brane 
in 5-dimensional dS Schwarzschild background. 
Similarly, we impose the following condition to obtain 
space-like brane metric instead of eq.(\ref{cd1}):
\bea
\label{cd3}
-e^{2\rho}\left( {\partial t \over \partial \tau} \right)^2
+e^{-2\rho}\left( {\partial r \over \partial \tau} \right)^2
= 1 \ .
\eea
Thus we get following FRW-like metric:
\footnote{This metric is also derived 
by Wick-rotation $\tau \to i\tau$ in eq.(\ref{met1}).}
\bea
ds^{2}_{4}=g_{ij} dx^i dx^j = d\tau ^2 +r^2 d \Omega_{3}^2 \ .
\eea
We again calculate the equations of motion
and the extrinsic curvature of space-like brane 
instead of (\ref{eom}) and (\ref{cd2}).  
These equations lead FRW like equation as follows:
\footnote{In ref.\cite{DF,DF1}, the similar equations to eqs.(\ref{HH}),
(\ref{H3}) were obtained in terms of Ricci scalar of the induced 
metric of the brane.} 
\bea
\label{H3}
H^2= -{1 \over l^2} +{1 \over r^2} - {\mu \over r^4} + 
{\kappa^2 \over 4} \ .
\eea
To cancel the cosmological constant, we take $\kappa ={2/l}$ 
in the same way of AdS Schwarzschild gravity\cite{SV}.  
We assume this equation can be rewritten by
using 4-dimensional energy $E_4$ and volume $V$ 
by the analogous form of the standard FRW equations:
\footnote{The reason why the sign of FRW equations
is different from the standard FRW equations (\ref{F1}) 
results from the condition (\ref{cd3}), namely 
$\tau \to i\tau$ in eq.(\ref{met1}).}
\bea
\label{F3}
H^2 &=& {1 \over r^2} - {8\pi G_4 \over 3}
{E_4 \over V}\ ,\quad E_4 ={ 3 \mu V  \over 8\pi G_4 r^4} \ . \\
\label{3FR1}
\dot H &=&  4\pi G_4 \left({E_4 \over V} 
+ p\right) - {1 \over  r^2}\ ,\quad p={\mu \over 8 \pi G_4 r^4 }\ .
\eea
Therefore we find the energy-momentum tensor is traceless
from eqs.(\ref{F3}) and (\ref{3FR1}) again.  

We stress again that {\it we can take cosmological constant to zero 
for FRW-like equation in space-like brane in
dS Schwarzschild background as the same way in the AdS/CFT 
correspondence.}  This will imply
that the dS/CFT correspondence can be valid for space-like
brane in dS Schwarzschild background.

%%%%%%%%%%%%%%%%%%%%%%%%%%%%%%%%%%%%%%%%%%%%%%%%%%%%%%%%%%%%%%%%
\section{The Cardy-Verlinde formula for the dS/CFT\\
correspondence}

In ref.\cite{EV}, E. Verlinde showed that the $d$-dimensional 
FRW equation can be regarded as an analogue of the
Cardy formula of 2-dimensional CFT\cite{CD}.  

\be
\label{CV1}
S_4 =2\pi \sqrt{
{c \over 6}\left(L_0 - {c \over 24}\right)}\ .
\ee
For time-like brane of 5-dimensional dS Schwarzschild 
background, identifying
\bea
\label{CV2}
{2\pi \over 3}\left( E_4 r + {\Lambda V r \over 8\pi G_{4} } 
\right) &\Rightarrow& 2\pi L_0 \ , \nn
{ V \over 8\pi G_4 r} &\Rightarrow& {c \over 24} \ ,\nn 
{ HV \over 2 G_4} &\Rightarrow&  S_4 \ ,
\eea
FRW equation (\ref{F1}) has the form (\ref{CV1}). 
The effect of the cosmological constant 
appears in Cardy formula.  We included contribution
of the cosmological constant in $L_0$ 
because it shifts the vacuum energy.  
This means the cosmological entropy bound 
which is discussed in ref. \cite{EV} should be changed.  
The Bekenstein bound\cite{EV} in 4-dimensions is 
\bea
S \le S_{B}, \quad  S_{B} \equiv {2\pi \over 3}Er\ .
\eea
Using eq.(\ref{CV2}), the Bekenstein 
entropy bound should be changed as follows:
\bea
S \le S_{B}, \quad  S_{B} \equiv {2\pi \over 3}\left( Er 
+ {\Lambda V r \over 8\pi G_{4} } \right) \ .
\eea
Then we find out that the effect of the cosmological
constant appears in the change of the Bekenstein 
entropy bound. 

For the case of space-like brane, 
identifying
\bea
\label{CV3}
{2\pi \over 3}\ E_4 r  &\Rightarrow& 2\pi L_0 \ , \nn
{ V \over 8\pi G_4 r} &\Rightarrow& {c \over 24} \ , \nn 
{ HV \over 2 G_4} &\Rightarrow&  S_4 ,
\eea
which is identical with AdS Schwarzschild case\cite{SV} exactly.  

For both cases, the moments when the brane crosses 
the horizon\footnote{The time-like brane can only cross
the black hole horizon, but the space-like brane can
cross the black hole and cosmological horizons.} $r=r_{H}$ which is derived from $e^{2\rho(r_{H})}=0$,  
the Hubble parameter in (\ref{HH}) becomes
\bea
\label{hor}
H = \pm {1\over l}\ .
\eea
Here the plus sign corresponds to the expanding brane
universe and the minus one to the contracting universe.  
We choose the expanding case.
Note that eq.(\ref{hor}) is the same form as the case of 
AdS Schwarzschild black hole\cite{SV,SO}.  Using eqs.(\ref{gg}),(\ref{CV2}), 
we obtain 4-dimensional entropy $S_4$ as follows:
\bea
\label{ent}
S_4 ={V \over 2 l G_4}={ V \over 4 G_5}\ .
\eea
This entropy is nothing but the Bekenstein-Hawking 
entropy of 5-dimensional dS black hole similar 
to AdS/CFT correspondence\cite{SV,SO}
\footnote{The black hole entropy 
of AdS background agrees with that of 
dS background.}.  

We now understand that discovered relation between FRW
equations and entropy formulas in ref.\cite{EV} can be also
applied to dS Schwarzschild background.  If we take
time-like brane which is the boundary
of dS Schwarzschild background, the 
cosmological constant of the brane appears in FRW equations. 
Therefore the effect of cosmological constant 
contributes to raising the the Bekenstein 
entropy bound.  But if we take space-like brane in
dS Schwarzschild background, we obtain the approximately 
same result of AdS Schwarzschild black hole\cite{SV,SO}.  
The difference between space-like brane in 
dS Schwarzschild background
and time-like brane in AdS Schwarzschild background is
the sign of FRW equations.  
When the brane crosses the horizon of dS Schwarzschild 
black hole, both for time-like and space-like brane, 
the entropy formula of the CFT exactly agrees with 
the black hole entropy of 5-dimensional 
dS background as it happens in 
the AdS/CFT correspondence.  This implies that 
the holographic principle holds true for 
dS Schwarzschild background.

\section*{Acknowledgements}

The author would like to thank S. Nojiri and S.D. Odintsov
for reading the manuscript and useful discussions.
The author wishes to thank Y. Hyakutake, A. Ishibashi 
for helpful discussions and advices.  
This work was supported in part by the Japan 
Society for the Promotion of Science under the Postdoctoral 
Research Programs.


\begin{thebibliography}{99}
\bibitem{AdS} J.M. Maldacena, 
{\it Adv.Theor.Math.Phys.} {\bf 2} (1998) 231, hep-th/9711200;
E. Witten, {\it Adv.Theor.Math.Phys.} {\bf 2} (1998) 
253, hep-th/9802150;
S. Gubser, I. Klebanov and A. Polyakov, {\it Phys.Lett.} 
{\bf B428} (1998) 105, hep-th/9802109;
O. Aharony, S. Gubser, J. Maldacena, H. Ooguri and Y. Oz,
{\it Phys.Repts.} {\bf 323} (2000) 183, hep-th/9905111.
\\ Further references are contained therein.
\bibitem{AS} A. Strominger, hep-th/0106113; hep-th/0110087;
For review, see M. Spradlin, A. Strominger, A. Volovich, hep-th/0110007. 
\bibitem{dS}
C.M. Hull, {\it JHEP} {\bf 9807}(1998) 021, hep-th/9806146;
E. Witten, hep-th/0106109, P.O. Mazur, E. Mottola, hep-th/0106151; 
S. Nojiri, S.D. Odintsov, {\it Phys.Lett.}{\bf B519} (2001)145, 
hep-th/0106191; hep-th/0107134; hep-th/0110064;
E. Silverstein, hep-th/0106209; D. Klemm, hep-th/0106247;
A. Chamblin, N. D. Lambert, hep-th/0107031; 
J. Bros, H. Epstein, U. Moschella, hep-th/0107091;
E. Halyo, hep-th/0107169; A. J. Tolley, N. Turok, hep-th/0108119; 
T. Shiromizu, D. Ida, T. Torii, hep-th/0109057; C.M. Hull, hep-th/0109213;
S. Cacciatori, D. Klemm, hep-th/0110031; B. McInnes, hep-th/0110062;
V. Balasubramanian, J. de Boer and D. Minic, hep-th/0110108;
Y.S. Myung, hep-th/0110123; Ulf. H. Danielsson, hep-th/0110265; 
R.-G. Cai, hep-th/0111093.
\bibitem{CI} R.-G. Cai, Y. S. Myung and Y.-Z. Zhang, hep-th/0110234. 
\bibitem{EV} E. Verlinde, hep-th/0008140.
\bibitem{CD} J.L. Cardy, {\it Nucl. Phys.}{\bf B270}(1986)967. 
\bibitem{RS} L. Randall and R. Sundrum, 
{\it Phys.Rev.Lett.} {\bf 83} (1999) 3370, hep-th/9905221; 
{\it Phys.Rev.Lett.}  {\bf 83} (1999)4690, hep-th/9906064. 
\bibitem{SV} I. Savonije and E. Verlinde, {\it Phys.Lett.}{\bf B507}(2001)305, hep-th/0102042.
\bibitem{others}
D. Kutasov and F. Larsen, 
{\it JHEP} {\bf 0101} (2001) 001, hep-th/0009244;
F.-L. Lin, {\it Phys.Rev.} {\bf D63} (2001) 064026, hep-th/0010127;
S. Nojiri and S.D. Odintsov, hep-th/0011115 to appear in 
{\it Int.J.Mod.Phys.}{\bf A}; hep-th/0103078 to appear in 
{\it Class.Quant.Grav.}; B. Wang, E. Abdalla and R.-K. Su, 
{\it Phys. Lett.} {\bf B503} (2001) 394, hep-th/0101073
D. Klemm, A. Petkou and G. Siopsis, 
{\it Nucl.Phys.} {\bf B601} (2001) 380, hep-th/0101076;
Y.S. Myung, hep-th/0102184;
R. Brustein, S. Foffa and G. Veneziano, 
{\it Phys.Lett.} {\bf B507} (2001) 270, hep-th/0101083;
R.-G. Cai, {\it Phys.Rev.} {\bf D63} (2001) 124018, hep-th/0102113;
R.-G. Cai and Y.-Z. Zhang, hep-th/0105214; 
A. Biswas and S. Mukherji, {\it JHEP} {\bf 0103} (2001) 046, 
hep-th/0102138;
D. Birmingham and S. Mokhtari, {\it Phys.Lett.} {\bf B508} 
(2001) 365, hep-th/0103108
D. Klemm, A. Petkou, G. Siopsis and D. Zanon, hep-th/0104141; 
D. Youm, {\it Mod.Phys.Lett.} {\bf A16} (2001) 1263, hep-th/0105036;
S. Nojiri, O. Obregon, S.D. Odintsov, H. Quevedo and M.P. Ryan, 
{\it Mod.Phys.Lett.} {\bf A16} (2001)1181, hep-th/0105052;
R.-G. Cai, Y.S. Myung and N. Ohta, hep-th/0105070;
B. Wang, E. Abdalla and R.-K. Su, hep-th/0106086;
L. Gappiello and W. Muck, hep-th/0107238; 
I. Brevik, S. D. Odintsov, gr-qc/0110105.
\bibitem{SO} S. Nojiri, S.D. Odintsov and S. Ogushi,
hep-th/0105117 to appear in {\it Int. J. Mod. Phys.}{\bf A};
hep-th/0108172 to appear in {\it Phys.Rev} {\bf D}. 
\bibitem{SD} S. Nojiri, S.D. Odintsov, hep-th/0107134. 
\bibitem{DF} P. Binetruy, C. Deffayet and D. Langlois,
{\it Nucl. Phys.} {\bf B}(2000)269, hep-th/9905012.
\bibitem{DF1} P. Binetruy, C. Deffayet, U. Ellwanger and D. Langlois,
{\it Phys. Lett.} {\bf B 477}(2000)285, hep-th/9910219, 
Further references are contained therein.
\end{thebibliography}
\end{document}